\documentstyle[11pt,paspconf]{article}
\def\etal   {{et~al.\/}}
\begin{document}

\title{Chemical Enrichment from Massive Stars in Starbursts}
\author{Henry A. Kobulnicky}
\affil{University of California Santa Cruz, Lick Observatory, 
Santa Cruz, CA 95064, USA}
\begin{abstract}

The warm ionized gas in low-mass, metal-poor starforming galaxies is
chemically homogeneous despite the prevalence of large H~II regions
which contain hundreds of evolved massive stars, supernovae, and
Wolf-Rayet stars with chemically-enriched winds.   Galaxies with large
Wolf-Rayet star content are chemically indistinguishable from other
vigorously star-forming galaxies.  Furthermore, no significant
localized chemical fluctuations are present in the vicinity of young
star clusters, despite large expected chemical yields of massive
stars.  An ad-hoc fine-tuning of the release, dispersal and mixing of
the massive star ejecta could give rise to the observed homogeneity, but a more
probable explanation is that fresh ejecta from massive stars reside in
a hard-to-observe hot or cold phase.  In any case, the observed
chemical homogeneity indicates that heavy elements which have already
{\it mixed} with the warm interstellar medium 
(thus accessible to optical spectroscopy) are homogeneously {\it
dispersed} over scales exceeding 1 kpc.  Mixing of fresh ejecta with
the surrounding warm ISM apparently requires longer than the lifetimes
of typical H~II regions ($>$10$^7$ yrs).  The lack of observed
localized chemical enrichments is consistent with a scenario whereby
freshly-synthesized metals from massive stars are expelled into the
halos of galaxies in a hot, 10$^6$ K phase by supernova-driven winds
before they cool and ``rain'' back down upon the galaxy, creating
gradual enrichments on spatial scales $>$1 kpc.

\end{abstract}

\keywords{}

\section{Introduction}

Starburst galaxies contain hundreds to thousands of the most massive O
type stars and their Wolf-Rayet descendants (see other articles in this
volume).  Massive stars are generally concentrated in clusters with
typical diameters of $\sim$10--20 pc.  The duration of the starburst
events in low-mass galaxies is thought to be 10 Myr or less, so that
the duty cycle of massive star formation over the lifetime of the
galaxy is small (Dohm-Palmer \etal\ 1998).  In normal galaxies, these
massive star clusters dominate the production of ionizing photons, the
generation of mechanical energy input to the ISM, and the
nucleosynthesis of $\alpha$-process elements (e.g., O, Ne, S, Si).
Understanding the fate of these heavy elements after they are ejected
by stellar winds and supernovae explosions is part of the larger
objective to trace the life cycles of matter in the universe.

The impact of massive star evolution on a host galaxy is best studied
in low-mass systems where the effect is most pronounced.  Low-mass
galaxies encompass a wide variety of objects with nomenclatures
including blue compact dwarf galaxies (BCDG), H~II galaxies, dwarf
elliptical galaxies (dE), irregular (Irr), and low surface brightness
(LSB) galaxies.  Because low-mass galaxies are also relatively
metal-deficient (Lequeux \etal\ 1979; Skillman 1989), the heavy element
yield of a given number of massive stars has a relatively large impact
on the chemical properties of the surrounding gas.  The chemical
composition of the stars and gas in low mass galaxies may be measured
by optical spectroscopy of emission lines from prominent H~II regions
(warm ionized gas), absorption lines in stellar atmospheres (stars), or
X-ray spectroscopy of emission lines from highly ionized atomic species
(hot diffuse gas).  Here I consider only abundance measurements of the
warm photo-ionized gas in and around H~II regions.  Since oxygen is the
most easily measured element in H~II regions, the terms ``abundance''
or ``metallicity'' should be read here as the ``abundance of oxygen
relative to hydrogen by number.''

Assessing the impact of massive stars on the chemical content of their
host galaxies entails two distinct issues:  1) Do galaxies with large
populations of evolved massive stars (i.e., Wolf-Rayet Galaxies) show,
as a group, signatures of chemical enrichment from the current
generation of star formation?  2) Are there {\it localized} chemical
fluctuations on spatial scales comparable to individual giant H~II
regions that might be due to heavy elements recently synthesized and
released by massive stars (e.g., ``self-enrichment''---Kunth \& Sargent
1986; ``local contamination'---Pagel, Terlevich, \& Melnick 1986)?
Since chemical abundances are like fossils that record the previous
star formation activity, both types of elemental variations contain
information about the star formation and gas dynamical history of the
host galaxy.

\section{Chemical Abundances in Wolf-Rayet Galaxies}

Galaxies with large populations of evolved massive stars in the
Wolf-Rayet phase are sometimes termed Wolf-Rayet Galaxies (see
contribution in this volume by P. Conti).  There are presently no
quantitative criteria by which objects are granted this distinction,
and it has now become clear that the Wolf-Rayet phase is a normal
episode in the evolution of starburst galaxies.  Wolf-Rayet stars are
observed in nearly all galaxies, including the Milky Way, the
Magellanic Clouds, and even the extremely metal-deficient I~Zw~18
(Izotov \etal\ 1997; Legrand \etal\ 1997).  Nevertheless, some galaxies
show prominent Wolf-Rayet features in their integrated optical spectra,
indicating a large fraction of stars in the Wolf-Rayet phase.  Some
authors postulate that the interstellar gas in such galaxies may become
``contaminated'' by nitrogen-rich and helium-rich winds of the evolved
stars (Pagel, Terlevich, \& Melnick 1986).  For this reason, such
galaxies are often omitted from studies seeking to use
the chemical properties of low-mass galaxies to measure the
primordial helium abundance and other fundamental cosmological parameters
(Pagel \etal\ 1992; Olive \& Steigman 1995).

I have collected from the literature and self-consistently analyzed the
oxygen, nitrogen, and helium abundances from nearly 100 low-mass
galaxies, as compiled in Kobulnicky \& Skillman (1996).  Figure~1 and
Figure~2 show the resulting chemical abundances.  Filled circles
distinguish galaxies with strong Wolf-Rayet features from galaxies
without strong Wolf-Rayet features (open circles).  A comparison of the
two galaxy samples reveals no significant difference between their mean
nitrogen abundances.  Likewise, the mean helium abundances for the two
samples are also consistent with one another.  This result shows that,
as a class, galaxies with strong Wolf-Rayet features from evolved
massive stellar populations do not exhibit anomalous chemical
properties compared to other star-forming galaxies.  This result does
not mean that evolved massive starbursts do not contribute heavy
elements to the surrounding interstellar medium;  we may only infer
that the presence of strong Wolf-Rayet features alone is not sufficient
to distinguish galaxies which may have recently experienced an episode
of chemical enrichment.

\begin{figure}
\plotfiddle{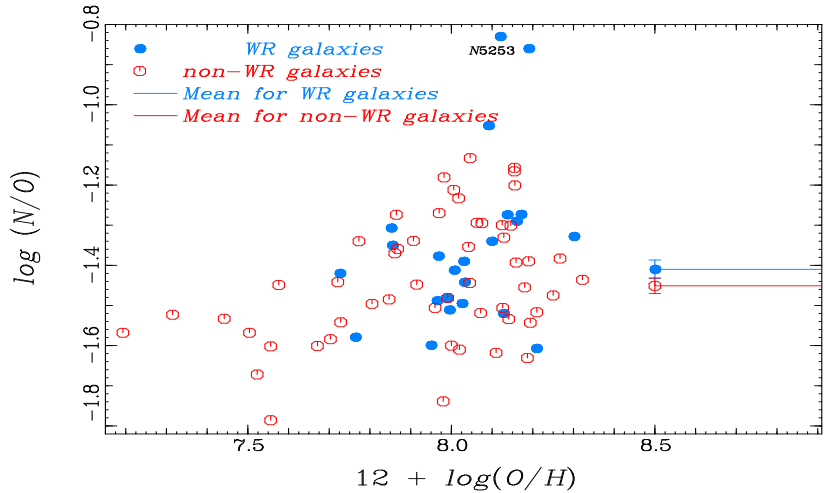}{65truemm}{-90}{50}{40}{-170}{225}
\caption{The oxygen abundance, as measured by 12+log(O/H),
versus the nitrogen-to-oxygen ratio in low-mass, metal-poor galaxies
from Kobulnicky \& Skillman (1996).
Filled circles distinguish galaxies with strong Wolf-Rayet features
from those without (open circles).  The mean N/O ratio for
Wolf-Rayet galaxies is consistent with that for non Wolf-Rayet galaxies.
There is no evidence that Wolf-Rayet galaxies are chemically
distinct in any way.}
\end{figure}

\begin{figure}
\plotfiddle{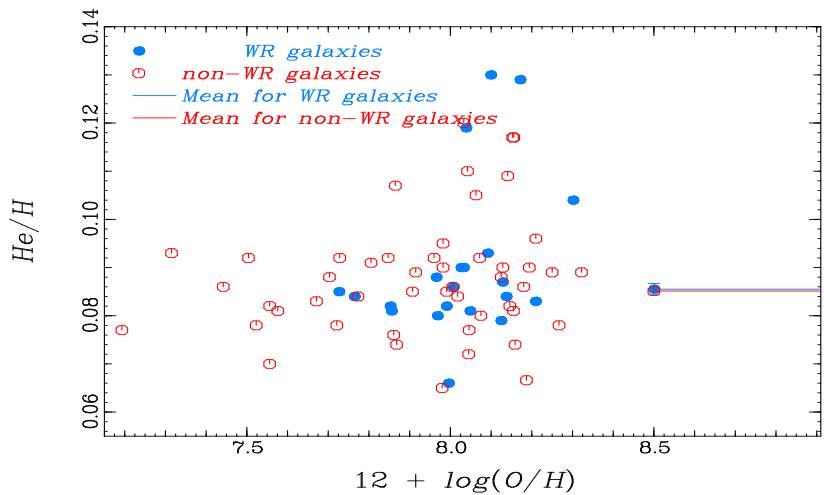}{65truemm}{-90}{50}{40}{-170}{225}
\caption{The oxygen abundance, as measured by 12+log(O/H),
versus the helium-to hydrogen ratio in low-mass, metal-poor galaxies
from Kobulnicky \& Skillman (1996).
Filled circles distinguish galaxies with strong Wolf-Rayet features.
from those without (open circles).  The mean He/H ratio for
Wolf-Rayet galaxies is consistent with that for non Wolf-Rayet galaxies.
There is no evidence that Wolf-Rayet galaxies are chemically
distinct in any way.}
\end{figure}

\section{Localized Chemical Abundance Fluctuations}

\subsection{Do Chemical Fluctuations Exist?}

\begin{figure}
\plotfiddle{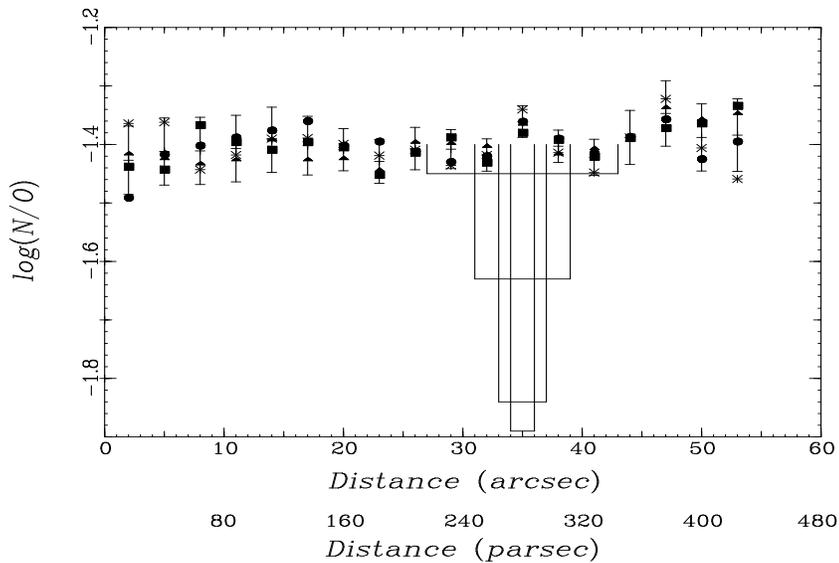}{65truemm}{-90}{50}{40}{-170}{225}
\caption{The N/O ratios along a 45$^{\prime\prime}$ strip of the ISM adjoining
cluster A in NGC~1569.  The expected magnitude
of chemical enrichment (predominantly O enrichment) is shown
by solid lines for four different spherical dispersal volumes.  The observed
variations are small in comparison to predicted enrichments, suggesting
some of the freshly-released elements are hidden in a hard-to-observe phase.}
\end{figure}

Clusters of massive stars are, in theory, capable of creating large
localized chemical enrichments (Esteban \& Peimbert 1995).  A typical
40 $M_\odot$ star may yield 2--6 $M_\odot$ of oxygen and several
$\times 0.1~M_\odot$ of nitrogen.  Individual Wolf-Rayet nebulae and
supernova remnants exhibit pronounced concentrations of fresh
nucleosynthesis products.  Yet, on the scale of giant H~II regions near
massive star clusters in NGC~4214 (Kobulnicky \& Skillman 1996) and
NGC~1569 (Kobulnicky \& Skillman 1997; Devost, Roy \& Drissen 1997),
NGC 2636 (Roy \etal\ 1996), and the SMC (Pagel \etal\ 1978; Russel \&
Dopita 1990) no measurable O, N, or He anomalies are seen in the
surrounding warm photoionized medium.   A quantitative comparison of
expected chemical pollution versus observed abundance fluctuations is
shown in Figure~3 for the case of the super star cluster A in
NGC~1569.  The measured N/O ratio along a 45\arcsec\ strip adjoining
the star cluster is plotted.  The N/O ratio is a particularly robust
measure of potential abundance fluctuations because it is relatively
insensitive to errors in the adopted electron temperature.  In
Figure~3, no substantial variations beyond the measurement
uncertainties are evident  despite a sensitivity to N or O yields of
just a few massive stars.  For example, the slightly-elevated N/O ratio
seen at the position number 12 at the 35$^{\prime\prime}$ mark could be
caused by as few as two 40 M$_\odot$ stars.

\begin{figure}
\plotfiddle{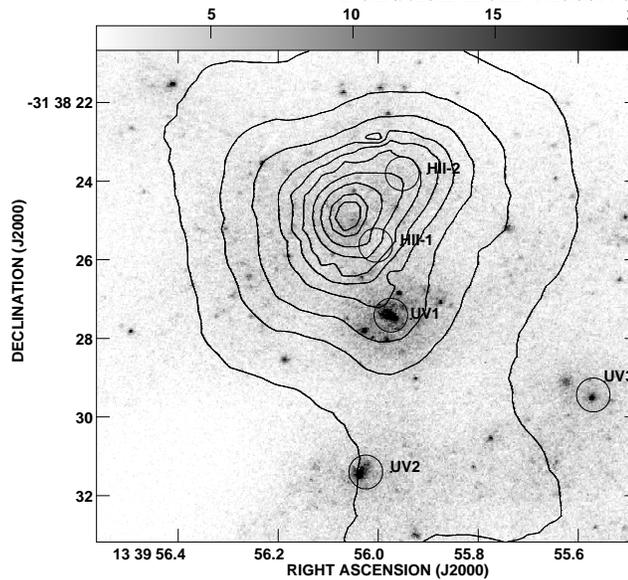}{65truemm}{0}{45}{45}{-170}{-80}
\caption{HST ultraviolet (greyscale) and H$\alpha$ (contours)
image of the central few arcsec of NGC~5253, showing the
3 UV-bright starclusters, and the location of HST Faint Object Spectrograph
apertures used to measure the region of nitrogen enhancement (Kobulnicky
\etal\ 1997).  The two apertures labeled H~II-1 and H~II-2 denote roughly
the position and size of the region where nitrogen appears overabundant by 
a factor of 2--3 compared to the rest of the galaxy.
}
\end{figure}

\begin{figure}
\plotfiddle{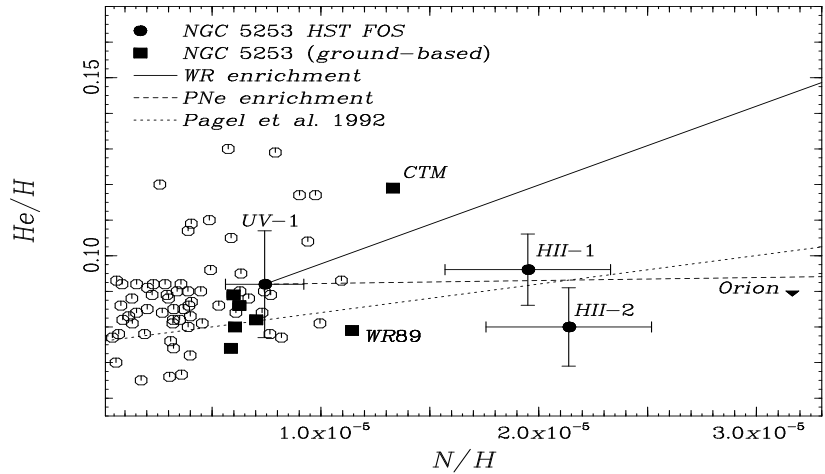}{60truemm}{-90}{50}{40}{-170}{215}
\caption{He/H versus N/H for 60 low--metallicity
galaxies and NGC~5253.   The ground-based observations of NGC~5253 
are shown with filled squares, and HST FOS data are plotted with 
filled circles. Positions H~II-1 and H~II-2 exhibit N abundances a 
factor of 3 above most metal--poor galaxies.  Lines
denoting the expected ratio of N and He enrichment from W-R star winds
and planetary nebulae are indicated.  
The lack of He enrichment accompanying the N enrichment
is inconsistent with pollution by Wolf-Rayet star winds. }
\end{figure}

NGC~5253 (Welch 1970; Walsh \& Roy 1989; Kobulnicky \etal\ 1997) and
possibly Markarian 996 (Thuan, Isotov, \& Lipovetsky 1996)  remain the
only galaxies where strong localized abundance anomalies are present.
NGC 5253 contains a central starburst region (Figure~4) overabundant in
nitrogen by a factor of 3 compared to the surrounding ISM.  The
nitrogen overabundant region appears to be $\sim$60 pc in diameter and
is centered on the heavily obscured starcluster to the NW of aperture
H~II-1.  Schaerer
\etal\ (1997) detect Wolf-Rayet features near this location. 
One hypothesis is that 6--10 Wolf Rayet stars have
``polluted'' this area with their nitrogen-rich winds.   The N
enrichment is equivalent to the nitrogen yield of $\sim$8 Wolf-Rayet
stars, or $\sim$120 Type I planetary nebulae.  However,  the lack of He
enrichment at the same locations (Kobulnicky \etal\ 1997) is difficult
to understand since localized enrichment by Wolf-Rayet winds should
produce elevations of both elements.  Ground-based spectra (Walsh \&
Roy 1989) do show an He overabundance along with the N overabundance,
lending credence to the Wolf-Rayet star scenario.

\subsection{Is Dilution the Solution to Pollution?}

Can the lack of observed enrichment be due to rapid dispersal and
dilution of the heavy elements?  Not likely.  Assuming an age of 10 Myr
for cluster A, homogeneous dispersal within a sphere of a given radius,
a filling factor for the ionized gas of 0.1, a gas density of 100
cm$^{-1}$, an IMF slope of -2.7 in the range 0.5---100 M$_\odot$ (see
Kobulnicky \& Skillman 1997 for details), the expected N/O deviations
for a variety of dispersal scales are shown with solid lines in
Figure~3.  Adopting any combination of inhomogeneous dispersal, higher
cluster age, lower filling factor or lower average gas density will
enhance the expected chemical variations.  Given typical expansion
speeds of supernova ejecta, the heavy elements {\it could } be
dispersed through a larger region than the largest simulated volume,
and thereby become undetectable.  However, such an extreme
rapid--dilution scenario requires a finely--tuned, ad hoc dispersal
mechanism to maintain the appearance of homogeneity of scales of
several hundred pc.  I~Zw~18 provides an extreme test of the rapid
dispersal hypothesis.  In a small metal-poor galaxy like I~Zw~18, even
two O stars can produce significant chemical enrichment.  Yet, both
starburst knots in I~Zw~18, separated by 300 pc, show identical
chemical properties (Skillman \& Kennicutt 1993).  It seems that the heavy elements produced by
the massive stars (down to 20 M$_\odot$) in cluster A 
in NGC~1569 are evidently not
seen in the warm ionized medium.

\section{Discussion}

On scales of 10 pc to over 1 kpc, low-mass galaxies appear chemically
homogeneous, indicating that the heavy elements are well dispersed
throughout the warm ionized phase of the interstellar medium.  The
heavy elements are not necessarily well-mixed, however, as
concentrations of massive star ejecta may exist on small, sub pc
scales, or they may be incorporated into a phase of the ISM other than
the warm, $\sim$10,000 K photoionized medium.  Given the lack of
visible chemical enrichment around major starbursts, several
explanations could be considered.

{\it 1) Perhaps the metals were
never produced/released in the first place.}  If the number of massive
stars originally present in the cluster has been overestimated based on
the remaining stellar content, then the expected mass of heavy elements
would be reduced.  An abnormally low upper mass cutoff in the IMF, or
an abnormally steep IMF could work to accomplish the observed effect.
Yet, NGC~4214 and NGC~1569 do contain Wolf-Rayet stars, so clearly the
most massive stars have been, and are still present in the bursts.  

{\it 2) An
alternative suggestion requires that black holes left by supernovae from
massive stars to engulf the metals produced by the progenitor (e.g., Maeder
1992).}  This idea merits further theoretical investigation, but unless
the lower mass limit for black hole formation, M$_{BH}$ is considerably
lower than $\sim$50 M$_\odot$, then the reduction of chemical yields
would be too minor to resolve this problem.  See discussion elsewhere in
this volume for current estimates of M$_{BH}$.

{\it 3) The last, and most probable explanation for missing metals requires
that the freshly-ejected metals reside in a hard-to-observe hot or cold
phase (e.g., Tenorio-Tagle 1996).}  Since supernovae and the superbubbles formed by concerted
supernovae contain copious X-ray emitting material, hot gas is the
preferred explanation.  The Cas A supernova remnant, for example,
contains between 4 M$_\odot$ (Vink, Kaastra, \& Bleeker 1996) and 15 M$_\odot$ 
(Jansen, Smith, \& Bleeker 1989) of X-ray emitting material, consistent with the amount of
ejecta expected from the progenitor.  The pending generation of 
X-ray observatories should be able to measure the mass and metallicity of hot
gas surrounding massive star clusters and make a direct comparison to
expectations based on starburst models.

\acknowledgments 
I thank the organizing committee for the invitation to present
this talk.  I am especially grateful to Peter Conti for
a timely 1992 colloquium at the University of Minnesota
which inspired much of this work.

\end{document}